\documentclass[preprint]{aastex}
\usepackage{psfig,emulateapj5}

\begin{document}

\slugcomment{To appear in ApJ Letters}

\title{Discovery of An Extremely Metal--Poor Galaxy: \\ Optical Spectroscopy of UGCA 292}

\author{Liese van~Zee}
\affil{Herzberg Institute of Astrophysics, National Research Council of Canada \\ 5071 W. 
Saanich Road, Victoria, BC V9E 2E7 Canada}
\email{Liese.vanZee@hia.nrc.ca}

\begin{abstract}
The results of optical spectroscopy of two \ion{H}{2} regions in UGCA 292 (CVn I dwA) are presented.
UGCA 292 is a nearby (D=3.1 Mpc) gas--rich dwarf irregular galaxy (M$_{\rm H}$/L$_{\rm B}$ $\sim$ 6.9) 
which was first discovered in a blind \ion{H}{1} survey.   The derived oxygen abundance
is the third lowest of known star--forming galaxies [12 + log(O/H) = 7.30 $\pm$ 0.05], making
UGCA 292  one of the nearest metal--poor galaxies known.
The derived N/O ratio is similar to that found in other low metallicity galaxies 
[log(N/O) = --1.47 $\pm$ 0.10], and is indicative of a primary origin for nitrogen.  
The derived oxygen abundance is consistent with closed--box chemical evolution
for this low mass galaxy.  The observed blue colors, high gas mass fraction,
and low metallicity suggest that UGCA 292 is relatively unevolved.    The possibility that future 
blind \ion{H}{1} surveys may yield similar low metallicity galaxies is discussed.
\end{abstract}


\section{Introduction}
In standard galaxy evolution models, the relative abundances of the elements increase 
as a function of time, as stars convert pristine (primordial) material into enriched material
\citep[e.g.,][]{TL78,CMG97,SHT97}.  It is thus of interest to identify extremely low metallicity 
galaxies at both high-- and low--redshift, since these galaxies may be ``young,''  
i.e., galaxies which have not processed much of their gas \citep[e.g.,][and references 
therein]{SS72,IT99,KO99}.  In addition, low metallicity galaxies provide a unique opportunity 
to determine the relative abundances of elements in pristine, unevolved, gas, and thus provide 
critical constraints on big bang nucleosynthesis \citep[e.g.,][]{PSTE92,MBF93,OSS97,Ie99}. 
However, despite concerted efforts over the last three decades to find nearby galaxies with
metallicities similar to I Zw 18 \citep[1/52th of solar\footnote{Adopting a solar oxygen 
abundance of 12 + log(O/H) = 8.93 (Anders \& Grevesse 1989).},][]{SS72,SK93}, 
only three other extremely low metallicity star--forming galaxies have been
found with 12 + log(O/H) $<$ 7.35: SBS 033-052 \citep[1/44th of solar,][]{Ie90,MHL92}, 
Leo A \citep[1/38th of solar,][]{SKH89,vSH00}, and HS 0822+3542 \citep[1/38th of solar,][]{Ke00}.
This {\sl Letter} reports the identification of another nearby metal--poor dwarf irregular galaxy, 
UGCA 292 (CVn I dwA).

UGCA 292 was observed as part of an on--going survey of elemental abundances
in nearby dwarf irregular galaxies \citep{vH00}.  
Many surveys for low metallicity galaxies target high surface brightness, 
``star bursting,'' dwarf galaxies, in part because their bright \ion{H}{2} regions 
make them easier to identify \citep[e.g.,][]{KS83,TMMMC91,MMC94,TIL95}.
An alternative approach, and the one adopted here,
 is to focus on extremely low luminosity galaxies, regardless of
their current star formation rate  \citep[e.g.,][]{SMTM88,SKH89,STM89}.
The drawback of this approach is that the \ion{H}{2} regions tend to be fainter, and thus
require significantly longer integration times (or larger aperture telescopes) to
reach an acceptable signal--to--noise ratio in the weaker lines.  
 Since the \ion{H}{2} regions tend to be faint, H$\alpha$ imaging observations are a 
necessary precursor to spectroscopic observations so that the slit can be 
positioned appropriately.  Thus, the present survey for low metallicity galaxies 
focusses on a subset of the isolated dwarf galaxy sample described in \citet{vZ00a},
since deep H$\alpha$ images were readily available.  
As one of the lowest luminosity galaxies in the imaging sample, UGCA 292
was an obvious target for spectroscopic observations.

UGCA 292 was first discovered by \citet{LS79}  during a search
for intergalactic \ion{H}{1} clouds in nearby groups of galaxies. 
UGCA 292 has an extremely low central surface brightness of 27.44 mag arcsec$^{-2}$
\citep{Me98} and is only marginally visible on the POSS plates. It is a member of
the Canes Venatici cloud, and a distance of 3.1 Mpc has been derived from photometry
of its brightest blue stars \citep{Me98}.  It has an apparent blue
magnitude of 16.10 (M$_{\rm B}$ = --11.43) and is extremely blue, with a 
luminosity weighted \bv~of 0.08 \citep{Me98}.  UGCA 292 has a current star formation rate
of 0.0019 M$_{\odot}$ yr$^{-1}$, which is typical of
the dwarf irregular class \citep{vZ00b}. With an integrated \ion{H}{1} flux of 17.60 Jy km s$^{-1}$,
corresponding to a neutral hydrogen mass of 4 $\times 10^7$
M$_{\odot}$ \citep{YvL00}, UGCA 292 is gas--rich.
 If UGCA 292 continued to form stars at the
present rate, it would take approximately 20 Gyr to deplete the current gas supply.
In addition, the derived M$_{\rm HI}$/L$_{\rm B}$  of 6.9 is unusually high for a dwarf galaxy, and
is comparable to DDO 154 \citep{He93} and \ion{H}{1} 1225+01 \citep{Se91}.

Given its low luminosity, blue colors, and relative gas--richness, UGCA 292 was one of the
best candidates in the H$\alpha$ imaging sample to be a low metallicity galaxy.
The results of optical spectroscopy of two \ion{H}{2} regions in this
low luminosity galaxy are presented in this {\it Letter}.   

\psfig{figure=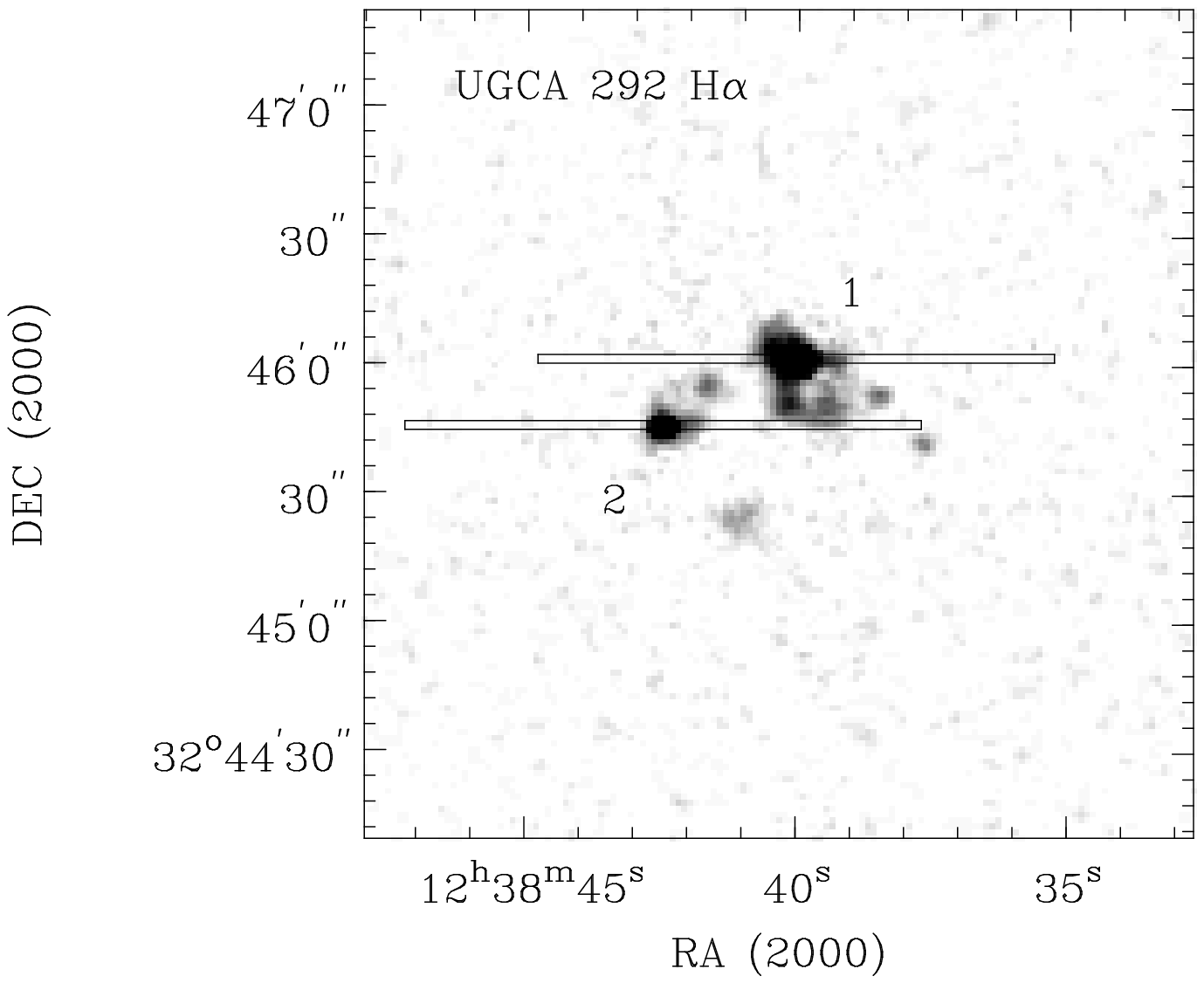,height=8.cm,angle=0,bbllx=50 pt,bblly=50 pt,bburx=500 pt,bbury=400 pt,clip=t}
\figcaption{H$\alpha$ image of UGCA 292.  The two HII regions are labelled, and
the
positions and orientations of the spectroscopic slits are shown.
 \label{fig:image}}

\section{Spectroscopic Observations}
Optical spectroscopy of two brightest \ion{H}{2} regions in UGCA 292 were obtained with the Double
Spectrograph on the 5m Palomar\footnote{Observations at the Palomar Observatory were
made as part of a continuing cooperative agreement between Cornell 
University and the California Institute of Technology.} 
telescope on 2000 January 31; the observations consisted of three 1200 second
exposures centered on each \ion{H}{2} region (Figure \ref{fig:image}).
  Since the \ion{H}{2} regions were not visible on 
the guider camera, the long slit (2\arcmin) was centered on a nearby star and then moved to 
the \ion{H}{2} region.
To minimize the effects of atmospheric differential refraction,
the observations were conducted near transit and the 2\arcsec~wide slit was oriented East--West to 
match the parallactic angle. 
The observations of UGCA 292--1 began shortly before transit; the airmass was
1.00 during all three exposures.  UGCA 292--2 was observed post--transit when
the parallactic angle was 83\arcdeg; the 
airmass increased from 1.01 to 1.07 during the observations.

Complete spectral coverage from 3600--7600 \AA~was obtained
by using a 5500 \AA~dichroic to split the light into two beams (blue and red).  The blue
camera was equipped with a 600 l/mm diffraction grating while the
red camera was equipped with a 316 l/mm grating.  Both sides of
the spectrograph were equipped with thinned 1024 $\times$ 1024 Tek CCDs, with gains of 2.0 e$^-$/ADU
and read noises of 8.6 e$^-$ (blue) and 7.5 e$^-$ (red).  The effective spectral resolutions
were well matched between the two sides, with a resolution of 5.0 \AA~(1.72 \AA/pix) on the
blue side and 7.9 \AA~(2.47 \AA/pix) on the red side.  The spatial scale of the long slit 
was 0.62\arcsec/pix on the blue and 0.48\arcsec/pix on the red side. 

The spectra were reduced and analyzed with the IRAF\footnote{IRAF 
is distributed by the National Optical Astronomy Observatories.} package.
The spectral reduction included bias subtraction, scattered light
corrections, and flat fielding with both twilight and dome flats.
The 2--dimensional images were rectified based on the arc lamp
observations and the trace of stars at different positions along 
the slit.   One dimensional spectra of the  \ion{H}{2} regions 
were extracted from the rectified images using a 3\arcsec~extraction
region (slightly larger than the seeing disk at the time of the observation).
Relative flux calibration was obtained by observations of standard stars from the 
list of \citet{O90}.  Since the night was non--photometric, only the standard 
stars observed contiguous to the  UGCA 292 observations were used to generate the 
sensitivity function. The optical spectra are shown in Figure \ref{fig:spec}.

\psfig{figure=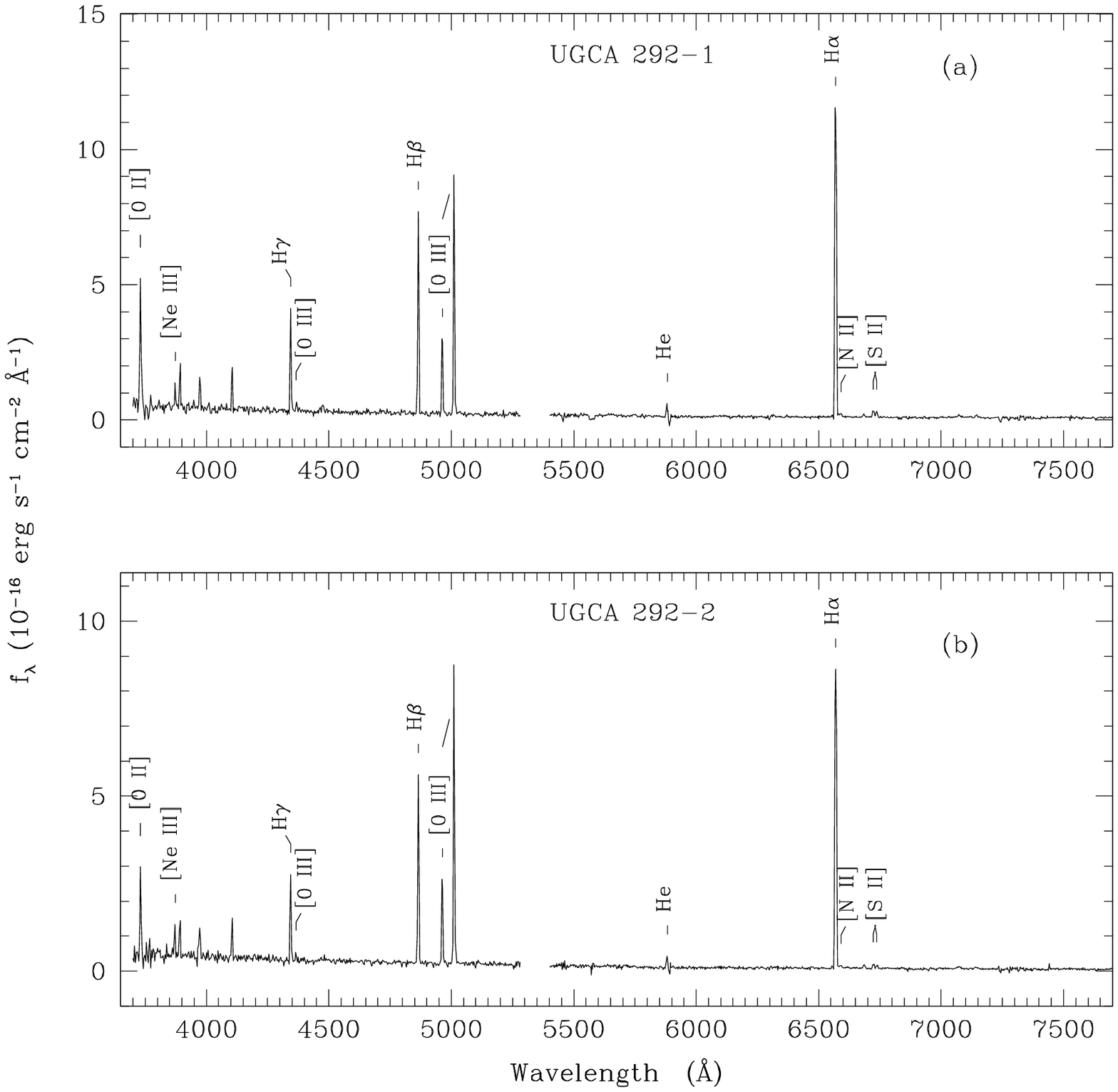,height=8.8cm,angle=0,bbllx=5 pt,bblly=150 pt,bburx=580 pt,bbury=700 pt,clip=t}
\figcaption{Optical spectra of \ion{H}{2} regions in UGCA 292;
 the major lines are marked.  Note the weakness of
[NII], which clearly indicates that these \ion{H}{2} regions are metal--poor.
 \label{fig:spec}}

\section{Results}

Derivation of oxygen and nitrogen abundances for the \ion{H}{2} regions
followed the methods described in \citet{vZ98}.  The reddening along the
line of sight to each \ion{H}{2} was derived from the observed line strengths
of the Balmer emission lines.  Since the derived reddening correction is
formally zero, the emission line ratios tabulated in Table \ref{tab:lines}
are the observed line ratios; however, the quoted errors in the emission
line ratios include the uncertainty associated with the reddening correction.  
In addition, while the line ratios for the higher order Balmer lines indicate
that there may be as much as 2 \AA~EW of underlying stellar absorption 
in UGCA 292--2, the tabulated values do not include a correction for
stellar absorption; application of such a correction has only a minor
effect on the derived oxygen and nitrogen abundances.

The temperature sensitive line [OIII] $\lambda$4363 was detected in both
\ion{H}{2} regions, so the electron temperature could be determined directly
using the emissivity coefficients from a version of the FIVEL program
\citep{DDH87}. The derived electron temperatures are high, but are similar
to those found in other low metallicity galaxies \citep[e.g.,][]{SK93,Se94,TIL95}.
The derived oxygen and nitrogen abundances for the \ion{H}{2} regions are listed in 
Table \ref{tab:lines}.  The derived oxygen abundances are similar for both
\ion{H}{2} regions, indicating that the overall oxygen abundance in UGCA 292 is
7.30 $\pm$ 0.05, or 1/42 of solar.  That is, UGCA 292 has an oxygen abundance
which is comparable to those of SBS 0335-052 \citep{IT99} and Leo A \citep{SKH89,vSH00},
and only slightly higher than that of I Zw 18 \citep{SK93}.   
In contrast to SBS 0335-052 and I Zw 18, both UGCA 292 and Leo A are  relatively
nearby galaxies with only moderate star formation rates.  

\psfig{figure=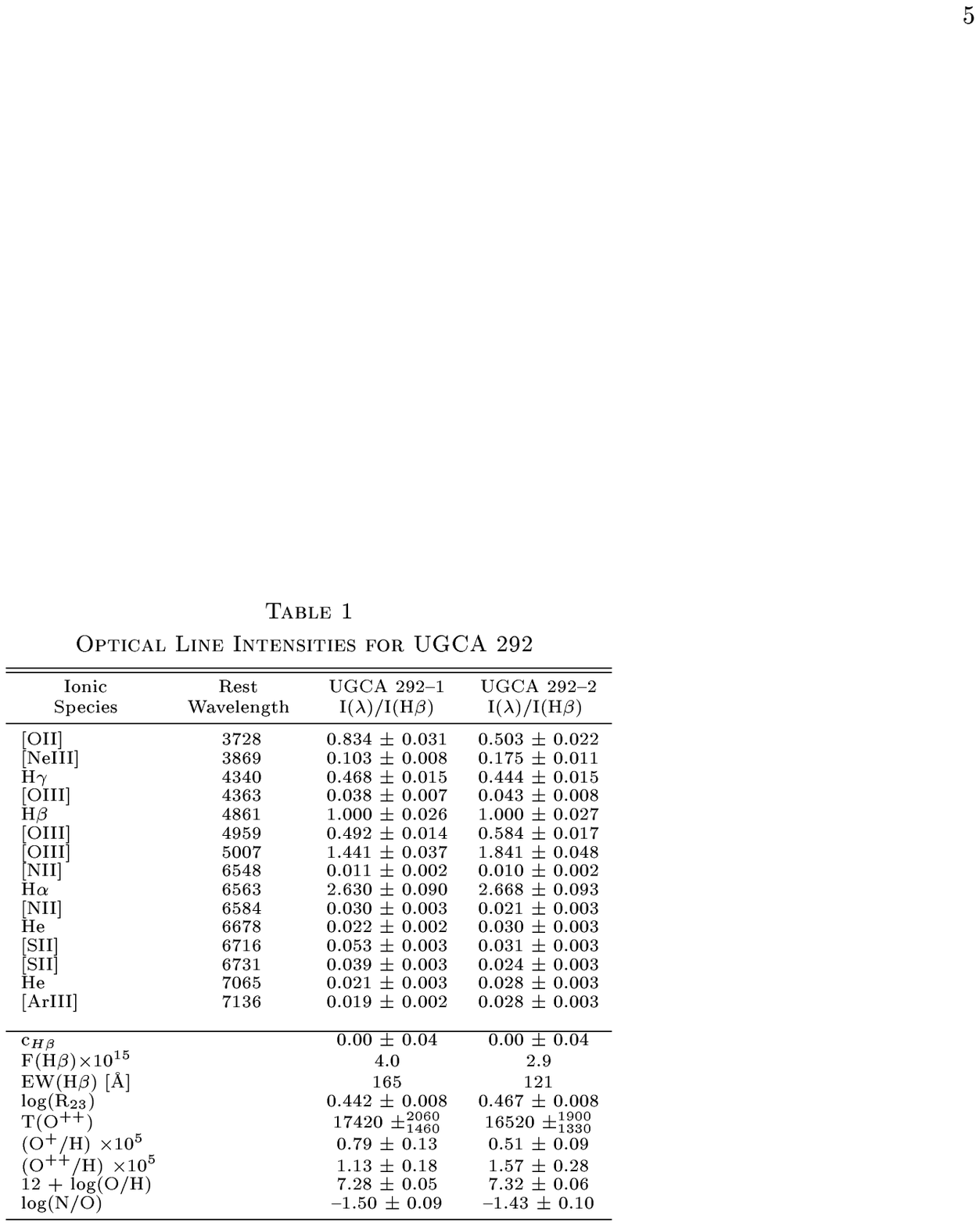,height=9.6cm,angle=0,bbllx=180 pt,bblly=250 pt,bburx=450 pt,bbury=520 pt,clip=t}

As one of the two nearest low metallicity galaxies, UGCA 292 provides a unique 
opportunity to investigate the evolutionary status of gas--rich, metal--poor galaxies 
in the local universe.  A fundamental issue is whether these galaxies represent
a population of ``young'' galaxies, which are undergoing their first episode of
star formation at the present epoch.  Based on the relative enrichment of
oxygen and nitrogen, \citet{IT99} argue that metal--poor galaxies are 
extremely young, with absolute ages less than 40 Myr.  
However, it is also possible that metal--poor galaxies have faint old stellar populations, 
but appear to be relatively unevolved because 
their previous star formation rates were extremely low \citep[e.g.,][]{Le00}.
Clearly, detailed star formation histories 
of  metal--poor galaxies are necessary to resolve this issue, but most of the 
starbursting galaxies are too distant to make such an analysis feasible.
However, the \ion{H}{2} regions in UGCA 292 have similar N/O abundances [log(N/O) = --1.46 $\pm$ 0.10] 
as the star bursting dwarf galaxies in \citet{IT99}, and UGCA 292
is close enough to us that its stellar population should be
well resolved by the {\it Hubble Space Telescope}.  Thus, future resolved stellar
photometry of UGCA 292 will provide an important test of the hypothesis that
all low metallicity galaxies are young.

An additional concern may be whether the observed elemental abundances in
metal--poor galaxies 
are representative of the chemical evolution which has taken place.
For instance, an extremely low mass galaxy may not retain its enriched materials
 if  supernovae ejecta  have sufficient kinetic energy to escape the shallow 
gravitational potential well \citep[e.g.,][]{DS86,MYTN97,STT98,MF99}.
In particular, mass loss of enriched material through supernovae ``blow--out'' may 
be important for starbursting galaxies, where multiple supernovae can 
significantly disrupt the interstellar medium \citep[e.g.,][]{M98,DB99}.  
On the other hand, for galaxies with more moderate
star formation activity, such as UGCA 292, mass loss is less likely to occur.

To determine if enriched gas outflow has occurred in UGCA 292, the expected
metallicity for a simple closed--box approximation was calculated from the
observed luminosity and gas content: 
\begin{equation}
Z = y~{\rm ln}~({1 \over \mu}),
\end{equation}
where $y$ is the elemental yield and $\mu$ is the gas mass fraction, derived from
the total baryonic mass.  Adopting a stellar mass--to--light
ratio of 0.35 and an oxygen yield of 2/3 solar, appropriate for a Salpeter IMF, the predicted closed--box 
abundance is 12 + log(O/H) = 7.28, in remarkable agreement with the observed abundance.
Thus, it appears that UGCA 292 has not lost a significant fraction of its enriched
material, and that the low oxygen abundance is representative of the chemical
evolution of the galaxy.

\section{Discussion}
Of a total sample of 25 low luminosity galaxies \citep{vHSB96,vHS97,vH00},
UGCA 292 was the only extremely low metallicity galaxy that was not previously identified
as such.  Other surveys of starbursting dwarf galaxies have shown how difficult it is
to find metal--poor galaxies \citep[e.g.,][]{TMMMC91}, so the low success rate of this survey
was not unexpected.  Nonetheless, it is discouraging to note that galaxies of lower
luminosity than UGCA 292 were included in the sample, yet no other extremely metal--poor galaxies were found.
In particular, if the luminosity--metallicity relation is valid
on all mass scales \citep[as indicated by][]{SKH89,ZKH,RM95},
low luminosity galaxies should have been ideal metal--poor candidates.
While a certain amount of scatter may be introduced into the luminosity--metallicity relation as a 
result of poorly determined distances,
it is interesting to note that, with the exception of UGCA 292, all of the extreme low luminosity 
galaxies observed for this project are relatively gas--poor, and their observed metallicities are
higher than  predicted from the luminosity--metallicity relation of \citet{SKH89}.  This
suggests that the apparent luminosity--metallicity relation 
may be a result of selection effects.
Most of the galaxies which have measured oxygen abundances are gas--rich, and
 are actively forming stars (i.e., \ion{H}{2} regions exist).
That is, aside from their luminosity (mass), their other physical parameters (gas mass fraction,
surface brightness, etc.) are very similar.  Galaxies which are gas--poor at the present time,
i.e., galaxies which have processed a large fraction of their materials, are usually excluded
from emission--line metallicity studies since it is very difficult to obtain an accurate oxygen abundance
for galaxies with faint (or non--existent) \ion{H}{2} regions.  With the exception of UGCA 292,
the other low luminosity galaxies observed for this project appear to have processed a large
fraction of their baryonic material, and thus may have enriched their interstellar medium
beyond that predicted by the gas--rich luminosity--metallicity relation.

Thus, an ideal survey for low metallicity galaxies should not only target low luminosity galaxies,
but should also require that the galaxies be gas--rich, in order to increase the probability
that they are relatively unevolved.  While it is unlikely that there are many more nearby previously
identified galaxies which are metal--poor (like UGCA 292), a viable approach may be to identify
new gas--rich galaxies from \ion{H}{1} surveys of the local universe.  There are several blind \ion{H}{1} surveys
currently underway that should yield interesting targets for future spectroscopic observations
\citep[e.g.,][]{ZBSS97,SS98,Be99,We99}.  The fact that UGCA 292 was first discovered
in such a survey is also suggestive that this may be an efficient means to identify
additional metal--poor galaxies.

\section{Conclusions}

 The results presented here, and those from other surveys, indicate that there are
a few galaxies in the local universe which have very low, almost primordial, oxygen
abundances.  The observed oxygen abundance in UGCA 292 is the third lowest of
known star forming galaxies and is consistent with a
simple closed--box chemical evolution model.  The observed N/O ratio is similar
to those found in other low metallicity galaxies, and is consistent with a 
primary origin for nitrogen in extreme low metallicity galaxies.
The identification of another metal--poor galaxy in the nearby universe suggests
that there may be several relatively unevolved galaxies at the present epoch.  However,
the question of whether some dwarf galaxies are ``young,'' undergoing their
first episodes of star formation, is still  open. A conclusive answer
to this question may be obtained from future observations
of the resolved stellar population in UGCA 292. 

\acknowledgements
Martha Haynes is gratefully acknowledged for providing access to the Palomar 5m for
these and other abundance measurements.  Elizabeth Barton, 
Evan Skillman, and Martha Haynes provided helpful comments on early versions of 
this paper.

\begin{table}
\dummytable\label{tab:lines}
\end{table}

\end{document}